\documentclass[10pt,a4paper,english]{article}
\usepackage[T1]{fontenc}
\usepackage[utf8x]{inputenc}
\usepackage{amsthm}
\usepackage{amsmath}
\usepackage{amssymb}
\usepackage{graphicx}
\usepackage{esint}

\makeatletter

\numberwithin{equation}{section}
\numberwithin{figure}{section}

\usepackage{graphics}\usepackage{epsfig}

\title{DARK NEUTRINOS}
\author{\Large{A. Nicolaidis} \bigskip \\
Theoretical Physics Department \\
University of Thessaloniki \\
54124 Thessaloniki, Greece \\
nicolaid@auth.gr}

\makeatother

\usepackage{babel}
\begin{document}

\date{}
\maketitle
\begin{abstract}
Solar, atmospheric and reactor neutrino experiments established that
neutrinos are massive. It is quite natural then to consider neutrinos
as candidate particles for explaining the dark matter in halos around
galaxies. We study the gravitational clustering of these neutrinos
within a model of a massive core and a surrounding spherical neutrino
halo. The neutrinos form a degenerate Fermi gas and a loaded polytropic
equation is established. We solve the equation and we obtain the neutrino
density in a galaxy, the size of the galaxy and the galactic rotational
curves. The available data favor a neutrino with a mass around 10eV.
The consequent cosmological implications are examined. 
\end{abstract}
\bigskip{}
\bigskip{}
\hspace{15pt}Solar, atmospheric, reactor and accelerator neutrinos
have provided compelling evidence for the existence of neutrino oscillations,
implying non-zero neutrino masses \cite{key-1}. Solar neutrino oscillations
indicate a mass squared difference $\Delta m_{SOL}^{2}=6\times10^{-5}eV^{2}$
while atmospheric neutrino oscillations suggest $\Delta m_{ATM}^{2}=2\times10^{-3}eV^{2}.$
Further, attributing the LSND anomaly to neutrino oscillations implies
the existence of a fourth neutrino, a sterile one, with a mass above
1 eV \cite{key-2}. Cosmology offers further insights on the neutrino
mass scales\cite{key-3}-\cite{key-6}. It is inevitable that neutrinos have
participated in gravitational clustering around massive galaxies,
constituting part of the dark matter. Our purpose is to investigate
the neutrino clustering effect, obtaining information on the cosmological
presence of massive neutrinos.

The universe today exhibits structure on many scales. Galaxies range
in mass, with most bright galaxies having masses of $10^{10}-10^{12}\, M_{\odot}$.
The sizes and distributions of present-day galaxies reflect the spectrum
of initial density fluctuations, seeded by random motion or cosmic
strings \cite{key-7}. We expect that each galaxy attracts the cosmological
neutrinos in its neighbourhood, creating a neutrino halo of an average
mass $M_{\nu}\approx3\times10^{10}\, M_{\odot}$.

Our idealized galaxy consists of a spherical massive core (of a mass
$M_{c}=\sigma M_{\odot}$ and a radius $r_{c}$ of few kpc) surrounded
by a spherical neutrino halo. Hydrostatic equilibrium prevails 
\begin{equation}
\frac{1}{r^{2}}\frac{d}{dr}(\frac{r^{2}}{mN}\frac{dP}{dr})=-4\pi GmN\label{eq-1}
\end{equation}
 where $N(r)$ is the neutrino density, m is the neutrino mass and
$P(r)$ is the pressure of the neutrino gas. Considering that the
neutrinos form a degenerate Fermi gas we obtain a polytrope rquation
\begin{equation}
P=(\frac{3}{4g\pi})^{2/3}\frac{h^{2}}{5m}N^{5/3}\label{eq-2}
\end{equation}
 with the degeneracy factor $g=2$ (left-handed neutrinos and right-handed
antineutrinos). Equs.(\ref{eq-1}) and (\ref{eq-2}) lead to 
\begin{eqnarray}
\frac{1}{r^{2}}\frac{d}{dr}\left(r^{2}\frac{1}{N^{1}/3}\frac{dN}{dr}\right)=-\gamma N\label{eq-3}\\
\gamma=m^{3}\left(3g^{2}\right)^{1/3}\left(4\pi\right)^{5/3}G/h^{2}
\end{eqnarray}
 Introducing the (dimensionless) variables p and x through 
\begin{equation}
N(r)=\frac{g}{6\pi^{2}}\left(\frac{2m^{2}}{\hslash^{2}}\right)^{3/2}\left(\frac{G\sigma M_{\odot}p}{r}\right)^{3/2}\label{eq-4}
\end{equation}
 
\begin{equation}
r=qx\label{eq-5}
\end{equation}
 
\begin{equation}
q=\left(\frac{3\pi\hslash^{3}}{4\sqrt{2}gm^{4}G^{3/2}M_{\odot}^{1/2}\sigma^{1/2}}\right)^{2/3}\label{eq-6}
\end{equation}
 equ.(\ref{eq-3}) becomes

\begin{equation}
\frac{d^{2}p}{dx^{2}}=-\frac{p^{3/2}}{\sqrt{x}}.\label{eq-7}
\end{equation}
 The variable p is related to the gravitational potential $V(r)$
through 
\begin{equation}
p=\frac{r(V_{0}-V)}{GM_{c}}\label{eq-8}
\end{equation}
 $V_{0}$ is the gravitational potemtial at the outer part of the
galaxy, where the density vanishes. Indeed the numerical solution
of equ.(\ref{eq-7}) always provides a finite $x_{0}$, where $p(x_{0})=0$.
The radius $R$ of the galaxy is then $R=qx_{0}$ (equ.(\ref{eq-5})).
The total mass of the galaxy ($M=M_{c}+M_{\nu}$) is related to the
derivative of $p(x)$ at $x_{0}$ by 
\begin{equation}
-\left(x\frac{dp}{dx}\right)_{x=x_{0}}=\frac{M_{c}+M_{\nu}}{M_{c}}\label{eq-9}
\end{equation}
 At $r=r_{c}$ the boundary condition is 
\begin{equation}
p(x=x_{c})=1.0\label{eq-10}
\end{equation}
 Equation (\ref{eq-7}) together with the boundary conditions (\ref{eq-9})
and (\ref{eq-10}), represents a loaded polytrope \cite{key-8}-\cite{key-11}.
Similar mathematical structures appear in the study of a galactic
nucleus, a neutron star, or in the Thomas-Fermi description of an
atom.

We solved numerically equ.(\ref{eq-7}), using the boundary condition
of equ.(\ref{eq-10}) and an arbitrary positive value for the derivative
of $p(x)$ at $x=x_{c}$. The numerical solution provides a function
$p(x)$ which rises up to a maximum, then decreases until it vanishes
at some point $x_{0}$ $(p(x_{0})=0)$. We evaluated also the derivative
of $p(x)$ in the neighbourhood of $x_{0}$ and subsequently the left-hand
side of equ.(\ref{eq-9}), thus determining the ratio $M_{\nu}/M_{c}$.
To obtain the desired ratio $M_{\nu}/M_{c}$, we repeat the numerical
evaluation with a different value for the derivative of $p(x)$ at
$x_{c}$, until we achieve the predetermined $M_{\nu}/M_{c}$ ratio.
The numerical study revealed the following essential features of the
loaded polytrope: 
\begin{enumerate}
\item The precise value of $x_{0}$ is largely independent of the ratio
$M_{\nu}/M_{c}$. More massive neutrino halos provide larger values
for the maximum of $p(x)$, but all neutrino densities vanish in the
vicinity of $x_{0}=2.0$. Thus the radius R of the galaxy is set up
by the constant q (equs.(\ref{eq-5}) and (\ref{eq-6})). 
\item The scale q, equ.(\ref{eq-6}), is very sensitive to the mass of neutrino
m. By increasing the mass of the neutrino by a factor 3, the size
of the galaxy is reduced by a factor 19. The overall data are well
reproduced with a neutrino mass at 10 eV. 
\item Equ.(\ref{eq-6}) gives then the numerical expression 
\begin{equation}
q=\frac{1.7}{\sigma^{1/3}}10^{5}\,\, Kpc\label{eq-11}
\end{equation}
 For a massive core $M_{c}=10^{11}\, M_{\odot}$, we obtain $q=36\, Kpc$
and therefore $R\approx70\, Kpc$, in agreement with the data for
the size of massive galaxies \cite{key-12}. It is impressive that a scale
q which is expressed in terms of fundamental constants, such as the
Planck constant, the neutrino mass and the Newton constant, reproduces
accurately the galactic sizes. 
\item Through equ.(\ref{eq-4}) we obtain back the neutrino density $N(r)$.
We observe that at small distances, $N(r)$ behaves as 
\begin{equation}
N(r)\sim\frac{1}{r^{3/2}}\qquad\qquad\textmd{at small r}\ \label{eq-12}
\end{equation}
 While the neutrino density diverges as $r\rightarrow0$, the total
neutrino mass $M_{\nu}$ remains finite.
\end{enumerate}
Fig. 1 shows the neutrino density as a function of the rescaled distance
x. The upper curve corresponds to $M_{c}=10^{11}M_{\odot}$, $M_{\nu}=10M_{c}$,
while the lower curve corresponds to $M_{c}=10^{11}M_{\odot}$, $M_{\nu}=M_{c}$.
Near the galactic core, the neutrino density is high as $10^{7}$
neutrinos/cm$^{3}$, and over the uniform density of the big-bang
cosmology, the gravitational clustering provides locally an increase
by a factor $10^{5}$.

The spherical neutrino halo up to a rescaled distance x, gives a mass
$\mu_{\nu}(x)$, where 
\begin{equation}
\mu_{\nu}(x)=\sigma M_{\odot}\int^{x}y^{1/2}[p(y)]^{3/2}dy.\label{eq-13}
\end{equation}
 Obviously $M_{\nu}=\mu_{\nu}(x_{0})$. The galactic rotational velocity,
due to the neutrino halo, is then 
\begin{equation}
u=\left(\frac{G\mu_{\nu}(r)}{r}\right)^{1/2}.\label{14}
\end{equation}
 Fig. 2 shows the galactic rotation curves (upper curve corresponds
to $M_{c}=10^{11}M_{\odot}$, $M_{\nu}=10M_{c}$, lower curve corresponds
to $M_{c}=10^{11}M_{\odot}$, $M_{\nu}=M_{c}$) and there is agreement
with the trends of the experimental data.

We may reach our findings by making appeal to an even simpler model\cite{key-13}.
A gravitational potential well of the size of the galactic halo is
filled up with neutrinos, which are treated like a Fermi-Dirac gas
at zero temperature. The total number of neutrinos is related to the
Fermi momentum, itself determined by the minimum kinetic energy required
for the neutrino to escape from the galaxy. It is found then that
the radius of the galaxy is $R=q3^{2/3}$, in full agreement with
our results.

Is it possible to detect the cosmological relic neutrinos? The gravitational
clustering effect we studied here, enhances the local densities, however
the average neutrino energy is very small and the corresponding weak
cross-section (of the order of $10^{-53}$ $cm^{2}$) renders detection
by conventional means highly unlikely \cite{key-14}.

The few eV mass scale for the sterile neutrino, that our study suggests,
fits nicely with the findings of the reactor and short-based neutrino
oscillation experiments\cite{key-2}. On the other hand the cosmological
verdict for a few eV-mass sterile neutrino is rather unclear. The
standard $\Lambda CDM$ framework provides the constraint $m\leq0.5$
eV. A modified though $\Lambda CDM$ cosmology, with additional radiation,
can accommodate a few eV neutrino\cite{key-15}. In another direction,
there is strong evidence for the existence of a 17 keV massive neutrino\cite{key-16}-\cite{key-17}.
The coexistence of distinct mass scales for the neutrinos, sub-eV,
eV and keV, indicates the complexity of neutrino physics. Clearly
we need further experimental information and novel theoretical insights
to decode the hidden dynamics.

In summary we presented a simplified model for the gravitational clustering
of massive neutrinos in the galaxies. There is no adjustable parameter
in our analysis and despite its simplicity the overall features of
galactic dynamics are reproduced.

\part*{Acknowledgements}

I would like to thank Nicos Vlachos for helping me to use the program
MATHEMATICA in the numerical work. This research was supported in
part by the Templeton Foundation and the EU program \textquotedbl{}Human
Capital and Mobility\textquotedbl{}.

\section*{Figures Caption}

Figure 1: The neutrino density in a spherical galaxy as a function
of the rescaled distance x. Upper curve corresponds to $M_{\nu}=10M_{c}$,
the lower curve corresponds to $M_{\nu}=M_{c}$. For a $M_{c}=10^{11}M_{\odot}$,
real distance is obtained by multiplying the dimensionless x by 36
Kpc. \medskip{}

\noindent Figure 2: The galactic rotational velocity, due to a spherical
neutrino halo, as a function of the rescaled distance x. Upper curve
corresponds to $M_{\nu}=10M_{c}$, lower to $M_{\nu}=M_{c}$. For
a $M_{c}=10^{11}M_{\odot}$, real distance is obtained by multiplying
the dimensionless x by 36 Kpc.

\begin{figure}
\begin{center}
 \includegraphics[width=1\textwidth,height=0.8\textheight,keepaspectratio]{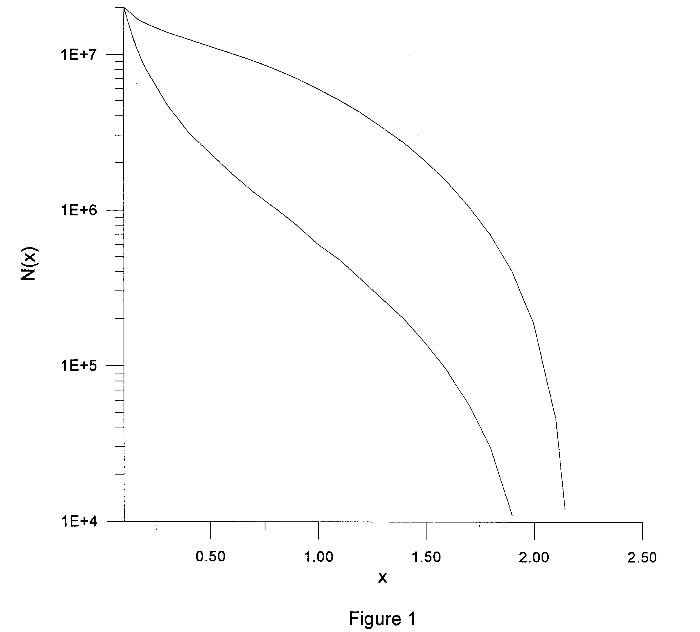}
\end{center}
\end{figure}

\begin{figure}
\begin{center}
 \includegraphics[width=1\textwidth,height=0.8\textheight,keepaspectratio]{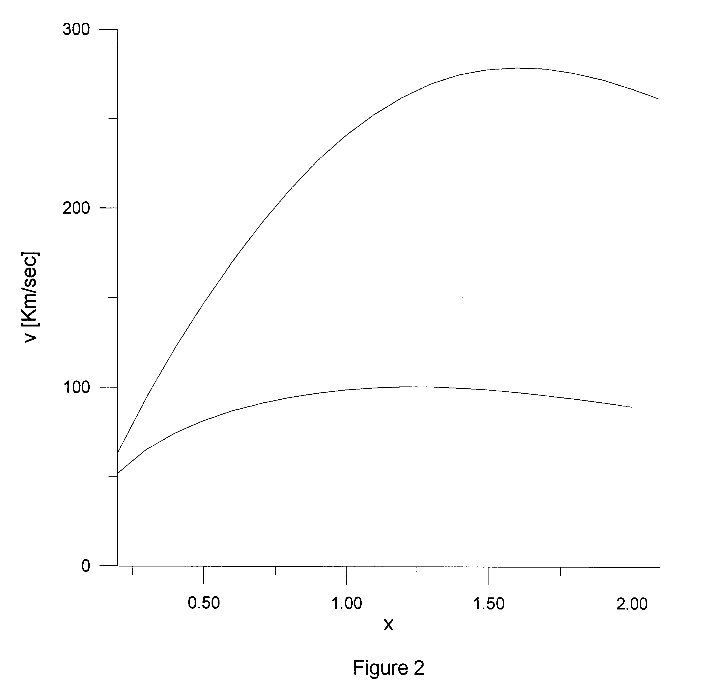}
\end{center}
\end{figure}

\end{document}